\pdfoutput=1
\documentclass[conference]{IEEEtran}
\IEEEoverridecommandlockouts

\usepackage{cite}
\usepackage{amsmath,amssymb,amsfonts}
\usepackage{graphicx}
\usepackage{textcomp}
\usepackage{xcolor}
\usepackage{booktabs}
\usepackage[hyphens]{url}
\usepackage{hyperref}
\hypersetup{pdftitle={Does Client Diversity Protect the Lightning Network? Evidence from the 2026 Core Lightning
Shutdown}, pdfauthor={Saurabh Kumar}}

\begin{document}

\title{Does Client Diversity Protect\\ the Lightning Network?\\
{\LARGE Evidence from the 2026 Core Lightning Shutdown}}

\author{
\IEEEauthorblockN{Saurabh Kumar}
\IEEEauthorblockA{Bitcoin Data Labs\\
sorukumar@bitcoindatalabs.org}
}

\maketitle
\thispagestyle{plain}
\pagestyle{plain}

\begin{abstract}
Client diversity is often cited as the Lightning Network's defence against a bug in any one implementation, yet
LND runs 80\% of public nodes (84\% counting one operator's fleet of about 2,000 tiny nodes) and holds 84\% of
public channel capacity, a capacity share that has barely moved since 2023.
We test what diversity buys with a natural experiment. On 26~August~2026, amid security fixes in every
implementation, Core Lightning (CLN) maintainers told operators who could not upgrade to disconnect their nodes.
Using 805 daily gossip snapshots, 132,360 on-chain channel closes and a client classifier validated against
self-identifying aliases and against the wallet that builds each node's funding transactions, we find that 22\% of
CLN nodes went dark overnight, ten times the usual rate, mostly Tor-reachable ones; a third were still dark
five weeks later. The rest of the network barely noticed. Reachability among all node pairs fell by 7.8
percentage points, a loss exceeded in 4 of 94 ordinary two-day windows and half that of a Tor denial-of-service
wave in July, while among CLN pairs it fell by 24.5 points, more than in any ordinary window. Calibrated failure
scenarios explain the containment: the blast radius depends on whether a failure reaches the hubs, not on how many
nodes run the failing software. Losing ACINQ's Eclair node alone costs the surviving network four times the reachability
lost with all 619 routable CLN nodes, and minority clients' reliance on LND for connectivity is what their size
implies; CLN nodes in fact peer with each other more than chance. Diversity statistics that count nodes mislead;
the quantity to track is which software the hubs run.
\end{abstract}


\section{Introduction}
\label{sec:intro}

The Lightning Network (LN)~\cite{poon2016bitcoin} routes Bitcoin payments over a network of two-party payment
channels. Its protocol is defined by the Basis of Lightning Technology (BOLT)
specifications~\cite{bolt2,bolt3,bolt7,bolt9}, and four independent implementations---LND (Lightning Network
Daemon), Core Lightning (CLN), Eclair and LDK (Lightning Development Kit)---interoperate on the same public
network. Client diversity is commonly invoked as a safety property: a bug in one implementation should not take
down the network. The public network does not look diverse. LND runs 80\% of public nodes (84\% if one operator's
fleet of about 2,000 tiny LND nodes is counted) and holds 84\% of public capacity, and its capacity share has
stayed between 84\% and 90\% for three years. Does the remaining diversity
protect anything, and is the concentration the risk it appears to be? We study the public (announced) graph, the
part of the network that forwards payments for others; wallets on unannounced channels attach to it through their
providers' nodes and are not visible in gossip.

Both questions have been hard to answer because client-correlated failures had not been observed. Topological
studies~\cite{rohrer2019discharged,seres2020topological,lin2020lightning,martinazzi2020evolving,lee2020robustness}
remove random or highly connected nodes from a snapshot; removing ``all LND nodes'' from a graph in which LND holds
85\% of capacity is trivially catastrophic and says little about realistic failures, which are partial,
version-scoped and temporary. In 2026 a natural experiment arrived. Between July and October every implementation
shipped security fixes, for different bugs, and on 26~August CLN maintainers told operators who could not upgrade
to take their nodes off the network. A sizeable cohort of one client went dark at once, in production. We measure
what happened and use it to calibrate a model of client-correlated failure.

The answer is that the outage hurt CLN's own users and almost nobody else. The nodes that complied were mostly
Tor-reachable nodes at the periphery of the graph, and the network as a whole lost about as much
reachability as in an unusual but ordinary two days. What determines the blast radius of a correlated failure is
whether it reaches the hubs: a single Eclair node, ACINQ's, matters more to everyone else than the entire CLN
cohort, and a transport-level failure (a denial-of-service wave against Tor in July) cost the network
twice as much reachability as the CLN shutdown. Counting nodes per client is the wrong way to assess diversity.

\paragraph*{Contributions}
\begin{itemize}
\item \textbf{C1: Validated client inference.} Feature-bit rules that stay valid on historical snapshots are
checked against 795 self-identifying aliases and against an independent on-chain signal, the wallet that builds a
node's channel funding transactions (\S\ref{sec:inference}).
\item \textbf{C2: Three years of concentration.} LND's share of public capacity stayed between 84\% and 90\% from
May~2023 to October~2026. Its share of nodes fell by five to seven points from late 2024, a decline that one
operator's fleet of small LND nodes hides in the raw counts (\S\ref{sec:concentration}).
\item \textbf{C3: An observed client-correlated outage.} We measure the CLN shutdown against churn baselines and a
control client: who complied, how long nodes stayed dark, how peers and the closing market reacted, and how the
event differs from transport failures and single-operator closing waves that resemble incidents
(\S\ref{sec:outage}).
\item \textbf{C4: What decides the blast radius.} Calibrated on C3, compared with 94 ordinary two-day windows and
with targeted baselines, and tested against a null model of how clients connect, client-correlated failures stay
contained; hubs, not cohorts, carry the risk (\S\ref{sec:resilience}).
\end{itemize}
All data are public gossip and blockchain data; code, aggregate datasets and every number quoted in the text are
released (see Data and Code Availability).

\section{Background and Related Work}
\label{sec:background}

\paragraph*{Gossip and channel state}
Nodes announce themselves (\texttt{node\_announcement}, including a feature-bit vector~\cite{bolt9}) and the
routing policy of each channel direction (\texttt{channel\_update})~\cite{bolt7}. A \texttt{channel\_update} can
mark its direction \emph{disabled}; implementations do so when the peer has been unreachable for some minutes.
The volume and propagation delay of such updates have been measured~\cite{gogge2023convergence}. A node that is
offline cannot announce anything; its \emph{peers} disable their direction toward it. Outages are therefore visible
as peer-side disabled flags, which is how we measure them. Node availability and reliability have also been
studied empirically~\cite{waugh2020availability}.

\paragraph*{Closing channels on-chain}
A channel is funded by a 2-of-2 output. A cooperative close spends it with a transaction both parties sign; a
unilateral (force) close broadcasts the latest commitment transaction, recognizable by its encoded commitment
number in \texttt{nLockTime} (upper byte \texttt{0x20}) and the input \texttt{nSequence} (upper byte
\texttt{0x80})~\cite{bolt3}. Its outputs are later swept (delayed \texttt{to\_local}, \texttt{to\_remote},
anchors, and the outputs of in-flight hash time-locked contracts, HTLCs) or, if a revoked state was published, claimed by the counterparty with a \emph{justice}
transaction. Prior work classified closes on-chain~\cite{grotschla2025lifecycle} and by implementation
pair~\cite{zabka2022empirical}, and linked on-chain and gossip data~\cite{kappos2021empirical,romiti2021crosslayer}.

\paragraph*{Resilience and attacks}
Rohrer et al.~\cite{rohrer2019discharged}, Seres et al.~\cite{seres2020topological}, Martinazzi and
Flori~\cite{martinazzi2020evolving} and Lee and Kim~\cite{lee2020robustness} found the LN robust to random
failures and fragile to the targeted removal of central nodes; Lin et al.~\cite{lin2020lightning} documented its
centralisation and showed that removing hubs splits it into many components. Attack studies exploit HTLC
handling, congestion, balance locking and correlated mass exits
\cite{harris2020flood,mizrahi2021congestion,perez2020lockdown,sguanci2023massexit}, and implementation defaults
shape the cost of some attacks~\cite{tikhomirov2020quantitative}. None conditions failure on the implementation or
observes one.

\paragraph*{Client inference}
Zabka et al.~\cite{zabka2021node,zabka2022empirical} infer implementations from default channel-policy and
node-announcement values; Espinasa-Vilarrasa et al.~\cite{espinasa2024decision} use BOLT~\#9 feature bits and
encrypted traffic metadata. We use feature bits, validate them against an on-chain signal, and keep the rules
valid across three years of feature-set changes.

\paragraph*{Monoculture and client diversity}
That software monoculture concentrates risk is a classic argument in security~\cite{geer2003cyberinsecurity}.
Birman and Schneider~\cite{birman2009monoculture} counter that the risk is often overstated: diversity multiplies
the distinct attacks available, and independently built implementations of one specification can share
vulnerabilities. In blockchains, client bugs have caused consensus failures, from Bitcoin's 2013 chain
fork~\cite{bip50} to Ethereum client incidents~\cite{prysm2023finality}, motivating differential testing across
clients~\cite{yang2021fluffy} and incentives for diversity~\cite{ron2026clientdiversity}. Lightning differs: a
client failure takes nodes offline rather than splitting consensus, so its cost depends on where in the payment
graph the failing nodes sit, which is what we measure.

\section{Data and Vantage Point}
\label{sec:data}

All inputs are truncated at a data cutoff of 4~October~2026.

\paragraph*{Daily gossip}
Our LND node dumps its view of the public graph (\texttt{describegraph}) daily. Between 8~May~2023 and the cutoff
(1,246 calendar days), 857 dumps produced usable extracts; the other days are missing, mostly in 2023 and early
2025, or were error responses from LND (``wallet locked'' after restarts). A quality pass excluded 51 partial dumps
(May--June~2024) and one stale day, leaving 805 days (65\% of the calendar). Coverage is complete enough for daily
analysis from May~2026, the window of \S\ref{sec:outage}. The node was replaced in mid-2024, which changed the
view: before, about 42k channels and 8.4k nodes with channels; after, 32--48k channels (median 41k) and 10--13k
nodes. We treat the two periods as \emph{vantage epochs} and compare levels only within an epoch. Gossip is a single
node's view of the public (announced) graph; unannounced channels, where much LDK (wallets) and Eclair (Phoenix)
usage lives, are not visible.

\paragraph*{On-chain closes and spends}
Our own Bitcoin Core node scans every block since January~2025 for spends of the 257,364 funding outputs of every
channel ever seen in our gossip. For each close we record the transaction, block time and type (cooperative, force,
splice); for each output of a force close, the transaction that spends it and its type from the witness script. The
data contain 132,360 closes (99,016 cooperative, 33,321 force, 23 splices) and 112,837 spends of force-close
outputs. Batched sweeps' fees are split across their inputs.

\paragraph*{Funding transactions}
For the 226,134 funding transactions of channels in our catalogue we record the funder wallet's construction
choices: \texttt{nVersion}, \texttt{nLockTime} policy, input \texttt{nSequence}, input and change script types,
output ordering, signature grinding, fee rate.

\section{Client Inference and Validation}
\label{sec:inference}

\paragraph*{Rules}
Table~\ref{tab:rules} lists the rules, applied in order. Implementation-specific feature bits of the node
announcement decide most nodes. Bits 2023 (script-enforced lease) and 31 (AMP, atomic multi-path payments), which LND
assigns outside BOLT~\#9, appear only on LND. Bit 37, or 29 together with 61, marks Eclair. CLN and Eclair share bit
11 together with 39 or 43; we separate them by bit 5, which every CLN release sets and no Eclair release sets by
default. Bits 39, 43 or 63 without 11 mark LDK. Fallbacks (LND's default colour and alias, and the modal default CLTV
delta, the number of blocks of CheckLockTimeVerify time-lock a node requires to forward a payment) yield low or
medium confidence; a node without announced features is never labeled LDK, because every LDK release announces
features. Operators with a publicly known implementation (ACINQ, Block) are attributed directly. Rules must be checked
against each implementation's defaults and against historical snapshots, not only the latest one: one earlier version
of our rules labeled ACINQ as CLN before 2025, when Eclair added bits 37 and 61, and another separated CLN from Eclair
by bit 55 (keysend), which CLN operators can switch off.

\begin{table}[t]
\centering
\caption{Client inference rules (v4), applied in order: the first that matches labels the node. Bits are
\texttt{node\_announcement} feature bits; CLTV is the modal \texttt{cltv\_expiry\_delta} of the node's own channel
updates.}
\label{tab:rules}
\footnotesize
\begin{tabular}{@{}rp{4.55cm}ll@{}}
\toprule
& Signal & Label & Confidence \\
\midrule
1 & Operator with a publicly known implementation: ACINQ (Eclair), Block's two routing nodes (LDK) & as listed &
operator \\
2 & Bit 2023 (script-enforced lease) or 31 (AMP) & LND & high \\
3 & Bit 37 (attribution data), or 29 (dual funding) with 61 (simple close); none of 51, 65, 561 & Eclair & high \\
4 & Bit 11 (extended gossip queries) with 39 (onion messages) or 43 (provide storage), and 5 (upfront shutdown
script) & CLN & high \\
5 & As rule 4, without bit 5 & Eclair & medium \\
6 & Bit 39, 43 or 63 (splicing), without 11 & LDK & medium$^{a}$ \\
7 & LND's default colour (\#3399ff) or alias (pubkey prefix) & LND & medium$^{b}$ \\
8 & CLTV default: 40 or 80 LND, 34 CLN, 144 Eclair, 72 LDK$^{c}$ & as listed & low \\
9 & None of the above & Unknown & low \\
\bottomrule
\end{tabular}
\par\vspace{2pt}\parbox{\columnwidth}{\footnotesize $^{a}$high with bit 7 and CLTV 72. $^{b}$low when the node
announces no features. $^{c}$CLTV 72 only for nodes that announce features.}
\end{table}

\paragraph*{On-chain signal}
A funding transaction is built by the funder's wallet. On channels whose two endpoints share a
high-confidence label (so the funder's software is known), two wallet families separate cleanly: LND's
btcwallet (\texttt{nLockTime}~0, \texttt{nSequence}~\texttt{0x0}) and Core/BDK-style wallets (anti-fee-sniping
\texttt{nLockTime} at the current height, \texttt{nSequence}~\texttt{0xfffffffd}), used by CLN, Eclair (via
bitcoind) and software based on the Bitcoin Development Kit (BDK). Since chain data do not reveal which endpoint funded a channel, we score
each node with a likelihood ratio in which a channel is funded by the node with probability $\theta=0.5$ and
otherwise by the peer, whose wallet distribution follows the peer's label; batch opens, where one node is
common to every channel funded by the transaction, identify the funder exactly. The signal measures the
\emph{wallet}, not the daemon: LND operators who fund channels from an external wallet look Core-style. It
separates LND from the other clients but cannot tell CLN from Eclair or identify LDK (our LDK sample is 99\% one
operator).

\begin{figure*}[t]
\centering
\includegraphics[width=\textwidth]{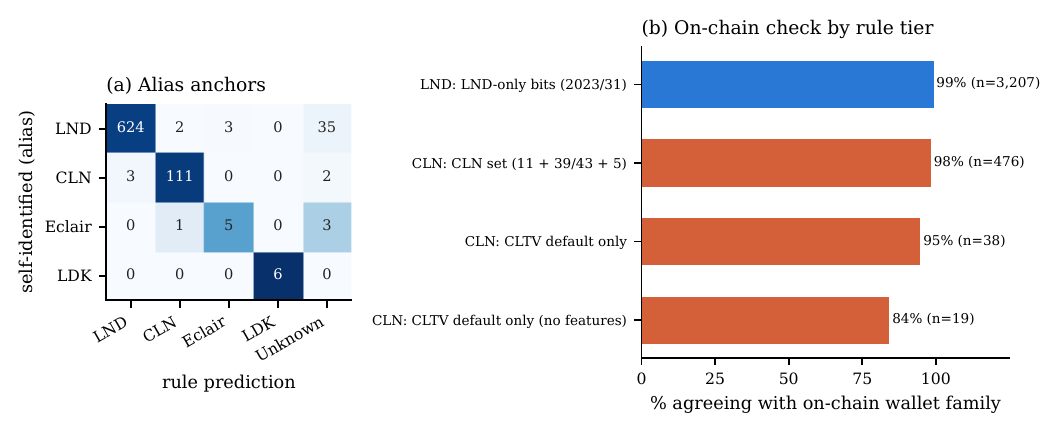}
\caption{Validation of client inference. (a) Nodes whose alias names their implementation vs.\ the rule
label. (b) Share of nodes in each rule tier whose channel-funding wallet agrees with the label (tiers with
$n\geq10$).}
\label{fig:validation}
\end{figure*}

\paragraph*{Validation}
Table~\ref{tab:classifier-eval} and Fig.~\ref{fig:validation} report two independent checks. Against 795 nodes
whose alias names their implementation (pooled over monthly snapshots; LND's default alias excluded because the
rules use it), precision and recall are 99.5\%/94\% for LND and 97\%/96\% for CLN. Most LND misses are nodes
without announced features, which the rules leave Unknown. Eclair (9 nodes, precision 62\%) and LDK (6) are too
few to grade; the three nodes wrongly labeled Eclair are LND nodes that only the CLTV fallback labels, so Eclair's
node count is an upper bound, although its capacity is dominated by ACINQ, which is attributed directly. Against the on-chain
wallet family (5-fold held out for high-confidence labels), nodes labeled by LND-only bits agree 99\% of the time
($n=3{,}207$) and the CLN set 98\% ($n=476$); the CLTV-only fallbacks are weakest (84--95\%) and account
for five of the nine alias disagreements (e.g.\ LND nodes of one node-in-a-box platform use CLTV~144, Eclair's
default). A learned baseline ($k$-nearest neighbours on the feature-bit vector) does not beat the rules on the same
held-out folds (macro-F1 0.92 vs.\ 0.93). Pooled over all high-confidence labels, held-out agreement between the
on-chain family and the label is 99.0\% on the 4,571 nodes where the chain decides, and 98\% of nodes with at least
five funding transactions get a decision; because the signal grades LND against everything else and most nodes are
LND, the per-tier figures above are the more informative ones.

\begin{table}[t]
\centering
\caption{Validation of client inference (rules v4-2026-10-14). (a) 795 nodes whose
alias names their implementation (pooled over monthly snapshots, evaluated on the latest snapshot where they have a
channel); ``Unknown'' counts nodes the rules leave unlabeled (mostly no announced features). A learned baseline
($k$-NN, Jaccard, on the feature-bit vector) vs.\ rules on the same held-out folds: macro-F1 0.92 vs.\ 0.93 ($n=761$). (b) Share of nodes in each rule tier whose channel-funding wallet agrees with the
label (LND btcwallet for LND, Core/BDK-style otherwise), nodes with decisive on-chain evidence; high tiers graded with
held-out evidence.}
\label{tab:classifier-eval}
\small
\begin{tabular}{lrrrr}
\toprule
\multicolumn{5}{l}{(a) Alias anchors} \\
Client & $n$ & Precision (\%) & Recall (\%) & Unknown \\
\midrule
LND & 664 & 99.5 & 94.0 & 35 \\
CLN & 116 & 97.4 & 95.7 & 2 \\
Eclair & 9 & 62.5 & 55.6 & 3 \\
LDK & 6 & 100.0 & 100.0 & 0 \\
\bottomrule
\end{tabular}

\vspace{4pt}
\begin{tabular}{llrr}
\toprule
\multicolumn{4}{l}{(b) On-chain wallet family} \\
Label & Rule tier & $n$ & Agree (\%) \\
\midrule
CLN & CLN set (11 + 39/43 + 5) & 476 & 98.3 \\
CLN & CLTV default only & 38 & 94.7 \\
CLN & CLTV default only (no features) & 19 & 84.2 \\
LND & LND-only bits (2023/31) & 3,207 & 99.2 \\
\bottomrule
\end{tabular}
\end{table}

\section{Concentration, 2023--2026}
\label{sec:concentration}

\begin{figure*}[t]
\centering
\includegraphics[width=\textwidth]{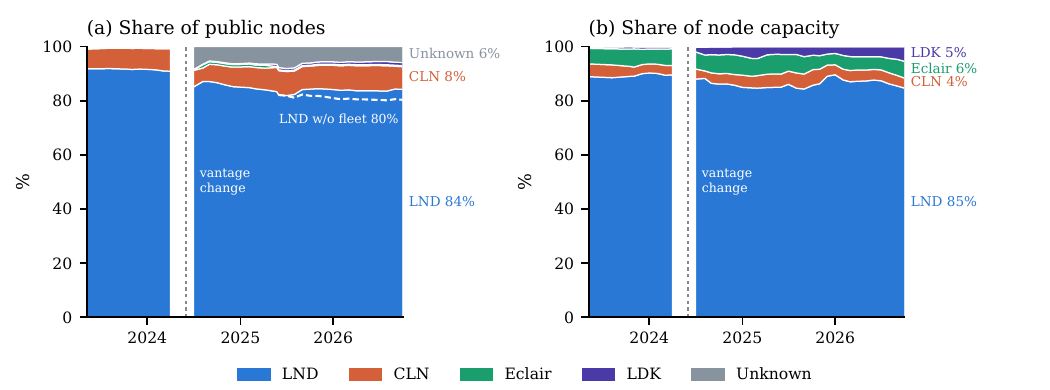}
\caption{Client shares of public nodes with at least one channel (a) and of node capacity (b), first gossip
snapshot of each month. The dashed line marks the change of our gossip vantage point (mid-2024); months with
unusable dumps are omitted. From mid-2025 the LND node share includes one operator's fleet of up to 1,980 small
nodes (\S\ref{sec:concentration}); the dashed line in (a) is LND's share of nodes without it.}
\label{fig:share}
\end{figure*}

Figure~\ref{fig:share} shows monthly client shares of public nodes and of node capacity (each channel counted
for both endpoints). LND's share of capacity stayed between 84\% and 90\% throughout. Its share of nodes fell
from 91.8\% (May~2023) to 84.2\% (October~2026). The step between epochs coincides with the change of our gossip
node and the appearance of many featureless (Unknown) nodes, not with a change in the network; within the first
epoch the share moved by about one point (91.8\% to 90.9\%). Within the second, the raw share looks stable (84.2\%
at the end against 85.1\% at the start), but one operator props it up. From July~2025 a managed fleet of small LND
nodes with aliases \texttt{LNT.<name>}, each with one or two channels, mostly to the operator's own hubs, grew to
1,980 nodes (March~2026), a fifth of all public nodes with channels; on the latest snapshot its 1,972 nodes hold
1.4~BTC (0.02\% of capacity). We count a node as
part of the fleet if its alias carries the prefix or at least half of its channels go to a fleet hub. Without
it, LND's share of nodes fell steadily (dashed line in Fig.~\ref{fig:share}(a)), from 85--87\% in the second half
of 2024 to 80.3\% in October~2026. In
capacity, the only clear within-epoch trend is LDK's share, from 1.7\% (July~2024) to 5.4\%, mostly Block's
routing nodes. On the latest snapshot (Table~\ref{tab:client-census}) LND runs 84.0\% of nodes, 80.1\% without
the fleet (77.5\% and 71.9\% counting only high-confidence labels), and holds 84.4\% of capacity; CLN runs 8.5\% of
nodes (10.6\% without the fleet) but holds 3.7\% of capacity; Eclair's 40 nodes hold 6.1\%.
Capacity is highly unequal within every cohort (Gini 0.92--0.97). The largest nodes are, if anything, more
concentrated by client than nodes in general: LND runs 90 of the 100 largest nodes of the routable core
(\S\ref{sec:resilience}) and 8 of the 10 largest; the other two are ACINQ's Eclair node and Block's LDK node.

\begin{table*}[t]
\centering
\caption{Client implementations of public Lightning nodes with at least one channel, gossip snapshot 2026-10-04 (rules v4-2026-10-14). Shares in \%; in parentheses, the share
counting only high-confidence labels (lower bound). W/o fleet: node share without the managed fleet of
Section~\ref{sec:concentration} (1,972 nodes, 1.4 BTC). Node capacity counts each channel for both
endpoints (total 7,390 BTC). Gini: inequality of node capacity within the cohort.}
\label{tab:client-census}
\small
\begin{tabular}{lrrrrrrrr}
\toprule
Client & Nodes & Node share & W/o fleet & Endpoints & Capacity (BTC) & Cap.\ share & Mean (BTC) & Gini \\
\midrule
LND & 8,290 & 84.0 (77.5) & 80.1 (71.9) & 85.6 & 6,237 & 84.4 (84.3) & 0.75 & 0.97 \\
CLN & 835 & 8.5 (7.2) & 10.6 (9.0) & 7.8 & 276 & 3.7 (3.7) & 0.33 & 0.92 \\
Eclair & 40 & 0.4 (0.1) & 0.5 (0.1) & 3.6 & 452 & 6.1 (5.6) & 11.31 & 0.95 \\
LDK & 115 & 1.2 (1.1) & 1.4 (1.3) & 1.9 & 421 & 5.7 (5.7) & 3.67 & 0.97 \\
Unknown & 588 & 6.0 (0.0) & 7.4 (0.0) & 1.2 & 4 & 0.1 (0.0) & 0.01 & 0.71 \\
\bottomrule
\end{tabular}
\end{table*}

\section{The 2026 Security Wave and the CLN Shutdown}
\label{sec:outage}

\paragraph*{Timeline}
On 19~July a researcher disclosed two memory-exhaustion denial-of-service bugs in CLN, already
fixed~\cite{pratap2026clndos}. On 7~August BTCPay Server reported that its versions before 2.4.2 let remote
attackers obtain the LND credentials (macaroons) of merchant nodes and that funds had been
stolen~\cite{btcpay2026advisory}; affected merchants found channels force-closed and balances
swept~\cite{rubboli2026btcpay}. This was an application-layer credential leak, not an LND protocol bug. On
11~August Lightning Labs published ten LND advisories, three of high severity, all for versions before
0.19~\cite{lightninglabs2026aug}. On 26~August CLN maintainers, after weeks of triaging a high volume of
vulnerability reports, advised operators to upgrade or, if they could not, to restart with \texttt{--offline},
which keeps the node running and watching the chain but drops every peer connection; the advice circulated during
the day and the project posted it publicly at 23:26~UTC~\cite{snoffline2026,coreln2026offline}. CLN 26.06.7
shipped on 28~August~\cite{cln26067}, LDK 0.2.6 on 9~September~\cite{ldk026} and Eclair 0.14.3 on
14~September~\cite{eclair0143}. Lightning Labs published eight more advisories (four for LND, four for btcd) on
21~September~\cite{lightninglabs2026sep}, CLN 26.06.8 followed on 22~September~\cite{cln26068}, and on 2~October
CLN warned that attackers were targeting nodes running 26.06.7 or earlier~\cite{coreln2026attack}. Every
implementation was affected; the bugs were different.

\paragraph*{Measure}
For client $c$ and day $t$, $D_c(t)$ is the share of channel directions pointing \emph{toward} nodes of $c$ that
their peers mark disabled, unweighted and weighted by capacity. A node restarted with \texttt{--offline} and a node
that was shut down look the same to its peers. Client labels are fixed per node on 1~August~2026, so the incident
cannot reshuffle cohorts. The baseline is 1--25~August.

\begin{figure*}[t]
\centering
\includegraphics[width=\textwidth]{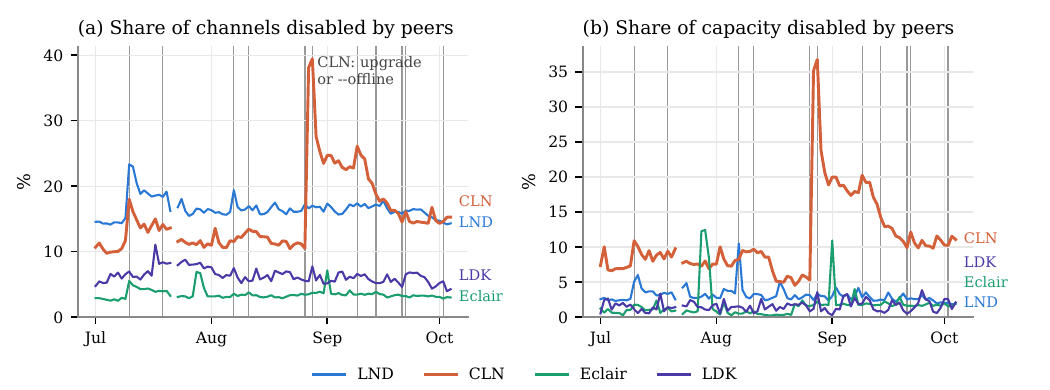}
\caption{Share of channel directions toward each client's nodes that peers mark disabled, daily,
July--October~2026; by channel count (a) and by capacity (b). Vertical lines: events of \S\ref{sec:outage}.}
\label{fig:outage}
\end{figure*}

\paragraph*{Magnitude}
On 27~August $D_{\mathrm{CLN}}$ rose from a baseline of $11.8\pm0.9\%$ to 38.1\% ($z=29$); by capacity from
7.4\% to 35.2\% (Table~\ref{tab:outage-effect}, Fig.~\ref{fig:outage}). LND, Eclair and LDK did not move
($|z|<2$). A difference-in-differences estimate against LND over the following two weeks, with a moving-block
bootstrap, gives $+14.2$ percentage points (95\% CI 11.2--16.0) by channel count and $+14.7$ (10.4--17.1) by
capacity. Removing, each day, the single node with the most disabled capacity leaves $+12.9$ (8.9--15.2), so
the capacity result is not one large node (Table~\ref{tab:outage-robustness}).

\begin{table}[t]
\centering
\caption{Share of channel directions toward each client cohort disabled by peers (\%). Baseline: daily mean $\pm$ sd, 2026-08-01 to 2026-08-25; peak: 2026-08-27; $z$ = (peak $-$ mean)/sd. Difference-in-differences, CLN vs.\ LND, 2026-08-27 to 2026-09-09 vs.\ baseline (95\% moving-block bootstrap CI, block 7 days): channels: +14.2 pp [+11.2, +16.0]; capacity: +14.7 pp [+10.4, +17.1].}
\label{tab:outage-effect}
\small
\begin{tabular}{lrrrrrr}
\toprule
 & \multicolumn{3}{c}{By channel count} & \multicolumn{3}{c}{By capacity} \\
\cmidrule(lr){2-4}\cmidrule(lr){5-7}
Client & Baseline & Peak & $z$ & Baseline & Peak & $z$ \\
\midrule
LND & 16.4 $\pm$ 0.8 & 16.6 & 0.3 & 3.5 $\pm$ 1.6 & 2.4 & -0.7 \\
CLN & 11.8 $\pm$ 0.9 & 38.1 & 29.4 & 7.4 $\pm$ 1.8 & 35.2 & 15.5 \\
Eclair & 3.2 $\pm$ 0.2 & 3.5 & 1.2 & 0.8 $\pm$ 0.6 & 1.7 & 1.5 \\
LDK & 6.3 $\pm$ 0.7 & 5.5 & -1.3 & 1.5 $\pm$ 0.6 & 1.2 & -0.5 \\
\bottomrule
\end{tabular}
\end{table}

\paragraph*{Who complied}
We call a node \emph{dark} when at least 80\% of the channel directions toward it are disabled, or when it is
absent from gossip. Of 624 CLN nodes not dark on 25~August, 137 (22.0\%) were dark on 27~August, holding 33\% of
CLN capacity. On an ordinary day the same transition (not dark on day $t-1$, dark on $t+1$) affects 2.3\% of CLN
nodes (median of 97 days, May--August; 95th percentile 5.6\%, maximum 12.3\%), so the call took about ten
times the usual number of CLN nodes off the network. LND nodes over the same days went dark at 6.4\%, within
their ordinary range (median 5.2\%, 95th percentile 9.5\%). Larger operators complied more: 32\% of the largest
capacity quintile against 8--27\% of the others (Table~\ref{tab:outage-compliance}). Compliance split by how nodes
are reachable: 28\% ($n=310$) of Tor-only CLN nodes went dark, 21\% ($n=150$) of dual-stack nodes and 7\% ($n=90$) of
clearnet-only nodes. Toward Tor-only CLN nodes the disabled share rose from 15.7\% to 39.0\%, toward dual-stack
nodes from 6.9\% to 47.4\%, while toward clearnet-only nodes it rose from 9.7\% to 19.0\% for two days and the
two-week difference-in-differences is zero (Table~\ref{tab:outage-robustness}). Tor-only LND nodes did not move
(25.6\% baseline, 23.4\% on 27~August), so this was not a Tor outage. A plausible reading, which our data cannot
confirm, is that home and node-in-a-box operators, typically on Tor, followed the call, and hosted operators
upgraded instead.

\begin{figure}[t]
\centering
\includegraphics[width=\columnwidth]{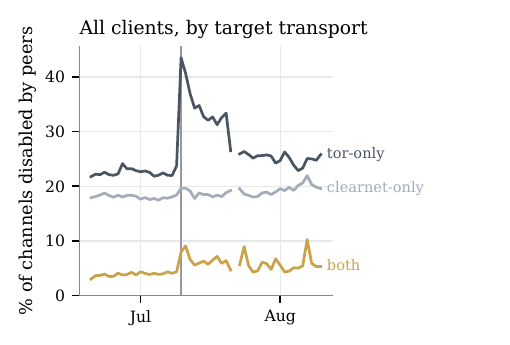}
\caption{Disabled share toward Tor-only, clearnet-only and dual-stack nodes, all clients pooled. On
10~July~2026, during a denial-of-service wave against Tor, Tor-only nodes of every client lost their peers at once.}
\label{fig:tor}
\end{figure}

\paragraph*{A transport-level contrast}
Gossip also shows a common-mode failure that was \emph{not} client-correlated. On 10~July the disabled share toward
Tor-only nodes of every client rose from 23.7\% to 43.5\% overnight (clearnet-only nodes flat) and decayed over two
weeks (Fig.~\ref{fig:tor}). It coincides with a denial-of-service wave against the Tor network, in which the number
of relays serving onion-service descriptors fell from 3,874 on 9~July to 1,780 on 12~July and recovered only after
22~July~\cite{aeo2026doswave,aeo2026hsdir,torcollector2026}; Tor-only Lightning nodes are reachable only through
those descriptors. Common-mode failures follow transport as well as software.

\begin{figure}[t]
\centering
\includegraphics[width=\columnwidth]{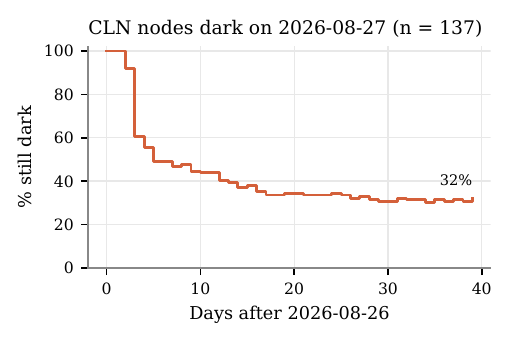}
\caption{Share of the CLN nodes that went dark after the shutdown call that were still dark, by day.}
\label{fig:recovery}
\end{figure}

\paragraph*{Recovery}
Of the 137 nodes, 39\% were back within three days and 53\% within a week; among nodes that returned, the median
time dark was three days (interquartile range 3--7), so most returns came shortly after the release of CLN 26.06.7
on 28~August, consistent with operators upgrading and reconnecting (gossip does not reveal versions). Twenty-two
percent never reappeared, even for a day, by 4~October; counting nodes that came back and went dark again, 32\% were
dark on that date (Fig.~\ref{fig:recovery}). The 2~October warning produced no comparable second wave
($D_{\mathrm{CLN}}$ 14.6\% on 2~October, 15.3\% on 4~October).

\paragraph*{Closing behaviour and confounds}
Read naively, closing data are full of ``incidents''. The two largest force-close days in the window were
single operators: on 3~August one LND node (NiceHash) was party to 226 force closes, and on 26~August---the day
of the CLN call---another LND node (Wallet of Satoshi), whose operator had announced a move away from its custodial
Lightning wallet~\cite{bitcoinworld2026wos}, was party to 205. The early-August cooperative-close wave was driven by
large operators closing hundreds of channels each: Kraken, which moved to a new node on
12~August~\cite{kraken2026node}, was party to 12.6\% of all closes on 2--16~August. We therefore separate
\emph{mass-close events}, a node party to at least 20 force closes in a UTC day (37 events since May, 30 by LND
nodes), from organic force closes. Organic force closes of channels involving a CLN node ran at 2.2 times their
July--August rate during 26~August--8~September, 2.6 times relative to channels between two LND nodes, whose rate
fell slightly (0.83 times; Table~\ref{tab:outage-compliance}). Within CLN, the channels of nodes that went dark were
force-closed at 70 per thousand channels in those two weeks, against 43 for CLN nodes that stayed up.
An \texttt{--offline} node can still broadcast its own commitment transactions, and chain data do not reveal which
side broadcast a close, so we cannot attribute these closes to peers. Force closes are not a sign of theft: we find
17 justice spends in 21 months, the last on 11~July~2026; a successful theft leaves none, so this bounds attempts
only from below.

\paragraph*{A contracting network}
The security wave coincided with the largest contraction of the public graph in our data, and the contraction was
not client-specific. Of the 38,081 channels in gossip on 1~August, 14,547 (38\%, holding 44\% of the capacity) were
gone by 4~October; 97\% of them closed on-chain, 10,464 in August, against 6,833 new channels. In the window of the
same length starting 1~May, 18\% of channels disappeared. Channels involving LND, CLN, Eclair and LDK nodes
disappeared at similar rates (38\%, 42\%, 32\% and 48\%); 79\% of the closes were cooperative, and ten operators,
led by Kraken's node migration, were party to 31\% of them. Exits by large operators, not client failures, account
for the network's shrinkage.

\paragraph*{Cost}
The median force close since May cost 1,222~sat in fees including the sweeps of its outputs, six times a
cooperative close (209~sat); all 8,787 force closes paid 0.17~BTC in total. The on-chain cost of the episode was
negligible; its cost was liquidity locked behind time-locks and channels lost.

\begin{table}[t]
\centering
\caption{(a) CLN nodes that went dark after the shutdown call (not dark on 2026-08-25, dark on 2026-08-27), by
pre-incident node capacity quintile. (b) Organic force closes per 1{,}000 open public channels per day involving
each cohort, 2026-07-01--08-25 (baseline) vs.\ 2026-08-26--09-08 (window); mass-close events (one node in
$\geq 20$ force closes in a UTC day) excluded.}
\label{tab:outage-compliance}
\small
\begin{tabular}{lrrr}
\toprule
\multicolumn{4}{l}{(a) Compliance by size} \\
Quintile & Capacity (BTC) & Nodes & Went dark (\%) \\
\midrule
Q1 & 0.000--0.004 & 125 & 23.2 \\
Q2 & 0.004--0.010 & 125 & 8.0 \\
Q3 & 0.010--0.041 & 124 & 19.4 \\
Q4 & 0.042--0.205 & 125 & 27.2 \\
Q5 & 0.208--44.488 & 125 & 32.0 \\
\midrule
\multicolumn{4}{l}{(b) Organic force-close rate} \\
\end{tabular}
\begin{tabular}{lrrrr}
Cohort & Baseline & Window & Ratio & vs.\ LND$\leftrightarrow$LND \\
\midrule
CLN & 1.40 & 3.07 & 2.18 & 2.63 \\
Eclair & 0.77 & 0.80 & 1.04 & 1.25 \\
LDK & 14.31 & 8.81 & 0.62 & 0.74 \\
LND$\leftrightarrow$LND & 1.02 & 0.85 & 0.83 & 1.00 \\
\bottomrule
\end{tabular}
\end{table}

\begin{table*}[t]
\centering
\caption{Robustness of the CLN outage effect. Baseline and peak as in Table~\ref{tab:outage-effect};
DiD vs.\ LND targets in the same subset, 95\% moving-block bootstrap CI. Target transport from node
announcements on 2026-08-01.}
\label{tab:outage-robustness}
\small
\begin{tabular}{lrrrr}
\toprule
Variant & Baseline (\%) & Peak (\%) & $z$ & DiD (pp) \\
\midrule
All targets, by channels & 11.8 & 38.1 & 29.4 & +14.2 [+11.2, +16.0] \\
All targets, by capacity & 7.4 & 35.2 & 15.5 & +14.7 [+10.4, +17.1] \\
By capacity, largest node removed & 6.1 & 31.8 & 20.6 & +12.9 [+8.9, +15.2] \\
Clearnet-only targets, by channels & 9.7 & 19.0 & 9.3 & -0.0 [-1.5, +1.6] \\
Tor-only targets, by channels & 15.7 & 39.0 & 9.4 & +23.1 [+20.8, +25.2] \\
Both targets, by channels & 6.9 & 47.4 & 39.6 & +16.1 [+10.0, +18.7] \\
\bottomrule
\end{tabular}
\end{table*}

\section{What Decides the Blast Radius}
\label{sec:resilience}

\paragraph*{Model}
We take the routable core of the public graph on 25~August~2026, the day before the call: channels whose two
directions are announced and enabled (73\% of announced channels; 4,435 nodes, of which 189 belong to the fleet
of \S\ref{sec:concentration}, almost all with a single channel). A payment of $a$ satoshis can
use a channel whose capacity is at least $a$ (optimistic) or $2a$ (balances split evenly). Failures are modelled as
\emph{exposure}, a fraction $p$ of a client's nodes being unavailable, not deletion of the client. Our main
measure is $R_{\mathrm{all}}(a)$, the share of the pre-failure node pairs still connected by a path of usable
channels, so pairs involving a failed node count as lost; we also report $R(a)$ among surviving nodes and
$R_c(a)$ among surviving pairs of client $c$. Connectivity is an upper bound on what routing achieves (no fees, no
in-flight HTLCs, no liquidity uncertainty beyond the threshold). Random scenarios average 50 draws. Eclair has
only 11 nodes in the core, so $R_{\mathrm{Eclair}}$ moves in large steps.

\begin{figure*}[t]
\centering
\includegraphics[width=\textwidth]{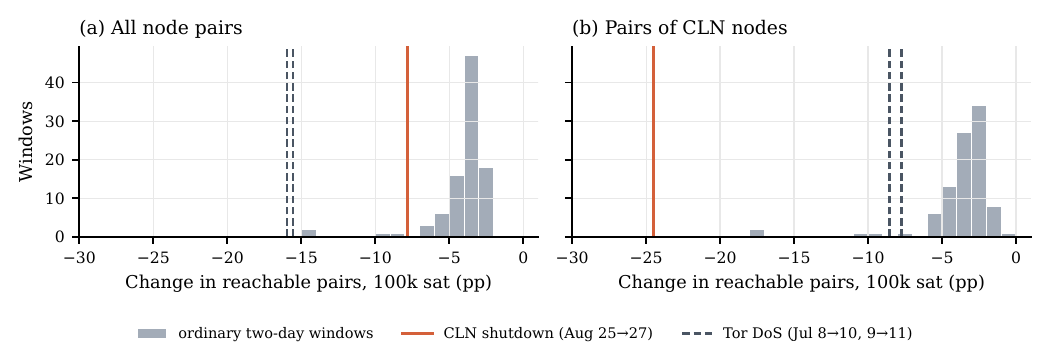}
\caption{Change in the share of pre-failure node pairs that can still route 100k sat, from each gossip snapshot to
the one two days later (nodes that lost every usable channel count as failed), May--August~2026. Grey: windows
without a known incident; solid line: 25$\rightarrow$27~August (CLN shutdown); dashed: 8$\rightarrow$10 and
9$\rightarrow$11~July (Tor denial-of-service wave). (a) All node pairs; (b) pairs of CLN nodes.}
\label{fig:placebo}
\end{figure*}

\paragraph*{The observed footprint against ordinary churn}
The observed footprint is the set of nodes of the 25~August core with no usable channel left on 27~August (383
nodes, 105 of them CLN). It reduced $R_{\mathrm{all}}$ for 100k-sat payments by 7.8 points. We compute the same
two-day footprint for every pair of snapshots two days apart from May to August (Fig.~\ref{fig:placebo}). In
94 windows without a known incident the loss had a median of 3.6 points; 4 of them lost more
than the CLN shutdown. For pairs of CLN nodes, counting pairs with a failed member as lost, the outage cost 24.5
points; no ordinary window cost more than 17.7 (median 3.2). The two windows spanning the Tor
event of 10~July cost the whole network 15.6--16.0 points, twice the CLN shutdown. The managed fleet does not drive
these comparisons. Its 189 nodes in the core, 157 of which are in the footprint, almost all have channels too small for
100k-sat payments, so they only enlarge the number of pairs; counting only pairs of nodes outside the fleet, the
outage cost 8.5 points against an ordinary median of 4.0, the same 4 windows lost more, and the Tor windows cost
17.3--17.8 points.

\begin{figure*}[t]
\centering
\includegraphics[width=\textwidth]{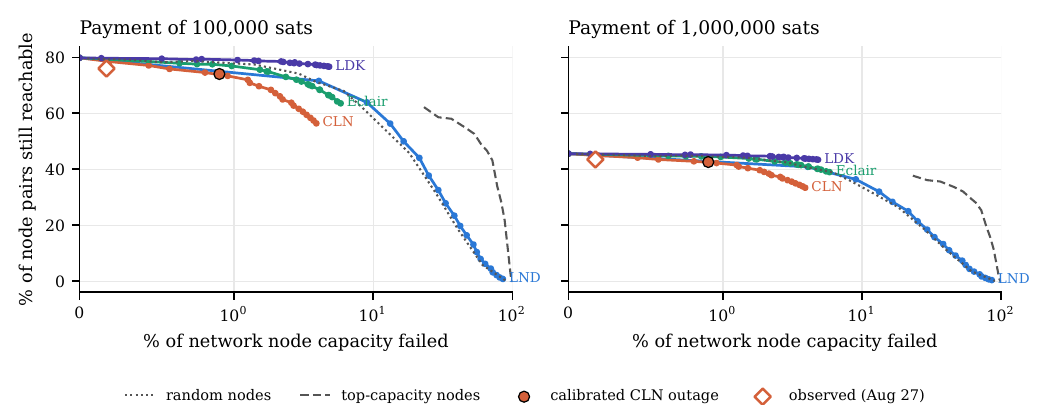}
\caption{Share of node pairs still reachable vs.\ share of network node capacity failed, for random failures
within each client (lines with markers), random failures across all nodes (dotted) and the largest nodes by
capacity (dashed). Dot: calibrated CLN outage; diamond: observed footprint of 27~August (CLN nodes). Routable core
of 25~August~2026; left: 100k sat, right: 1M sat payments.}
\label{fig:curves}
\end{figure*}

\paragraph*{Scenarios}
Table~\ref{tab:scenarios} and Fig.~\ref{fig:curves} compare failures. The calibrated scenario ($\hat p=0.22$ of
CLN nodes failing at random, 136 nodes) leaves $R$ among survivors at 78.8\%, against 78.2\% when the same number
of nodes fails at random across all clients: for the rest of the network, a CLN-wide outage behaves like random
failure, because CLN nodes are mostly peripheral. Drawing the failures with the observed compliance of each size
quintile and transport class instead gives the same result ($R=78.7\%$, $R_{\mathrm{all}}=73.9\%$). The damage stays inside the cohort: counting
pairs with a failed member as lost, CLN pairs fall from 79\% to 46\% connected, while among surviving CLN nodes
$R_{\mathrm{CLN}}$ is 76\%. Removing the same number of nodes chosen by degree leaves 27.6\%, by capacity 33.7\%.
Losing all 619 CLN nodes of the core leaves survivors at 76.2\%; losing its 11 Eclair nodes, which hold 5.8\% of
capacity and are dominated by ACINQ, leaves 63.9\%. Per unit of failed capacity, failures confined to Eclair do the
most damage to the surviving network and failures confined to LDK the least; counting pairs with a failed member as
lost (Fig.~\ref{fig:curves}), CLN's curve falls fastest, because its failed nodes are many and small.

\begin{table*}[t]
\centering
\caption{Failure scenarios on the 2026-08-25 public graph (channels enabled in both
directions), payment size 100,000 sats, optimistic liquidity. $R_{\mathrm{all}}$: \% of pre-failure node pairs still connected (pairs with a failed
node count as lost); $R$: \% of online node pairs connected by a path of
channels that can carry the payment; $R_c$: same as $R$, for pairs of surviving nodes of client $c$; LCC cap.: \% of usable capacity in the largest
component. Random scenarios: mean of the Monte Carlo runs.}
\label{tab:scenarios}
\small
\begin{tabular}{lrrrrrrrr}
\toprule
Scenario & Failed & Cap.\ (\%) & $R_{\mathrm{all}}$ & $R$ & $R_{\mathrm{CLN}}$ & $R_{\mathrm{Eclair}}$ & $R_{\mathrm{LDK}}$ & LCC cap. \\
\midrule
Intact & 0 & 0.0 & 79.8 & 79.8 & 79.2 & 81.8 & 79.1 & 100.0 \\
Observed: lost all usable channels, Aug 27 & 383 & 0.5 & 72.0 & 86.3 & 79.4 & 81.8 & 80.9 & 99.0 \\
  of which CLN nodes only & 105 & 0.2 & 76.0 & 79.8 & 79.4 & 81.8 & 79.1 & 99.6 \\
Calibrated CLN ($\hat p$ = 0.22) & 136 & 0.9 & 74.1 & 78.8 & 76.3 & 81.8 & 78.2 & 98.2 \\
Same count, random nodes & 136 & 3.1 & 73.4 & 78.2 & 77.0 & 79.8 & 77.1 & 93.9 \\
Same count, top degree & 136 & 70.5 & 25.9 & 27.6 & 24.8 & 27.8 & 17.8 & 8.9 \\
Same count, top capacity & 136 & 80.2 & 31.7 & 33.7 & 30.1 & 14.3 & 12.3 & 6.6 \\
All LND nodes fail & 3,704 & 85.4 & 0.8 & 28.3 & 31.8 & 50.9 & 12.2 & 2.3 \\
All CLN nodes fail & 619 & 3.9 & 56.4 & 76.2 & -- & 81.8 & 74.8 & 92.8 \\
All Eclair nodes fail & 11 & 5.8 & 63.6 & 63.9 & 59.1 & -- & 68.6 & 88.5 \\
All LDK nodes fail & 82 & 4.8 & 76.7 & 79.6 & 78.9 & 81.8 & -- & 91.1 \\
\bottomrule
\end{tabular}
\end{table*}

\paragraph*{Hubs, not cohorts}
Single nodes explain the gap between cohorts. Removing each of the 300 largest nodes of the core on its own, the
most critical by far is ACINQ's Eclair node: it holds 4.9\% of node capacity, and its loss alone cuts $R$ by 15.9
points, four times the loss from all 619 CLN nodes (3.7 points). The next most critical nodes, a Lightning service
provider and a payment processor running LND, cost 2.8 and 1.9 points; the 20 most critical nodes, 19 of them LND,
cost a median of 0.6 points each. Criticality follows a node's role rather than its size: the two largest nodes by
capacity, both exchanges, are not among the six most critical, whereas nodes that are the only adequate link for
many small nodes are. The software of these few nodes, not the client shares of Table~\ref{tab:client-census},
decides how far a client-correlated failure reaches.

\begin{figure}[t]
\centering
\includegraphics[width=\columnwidth]{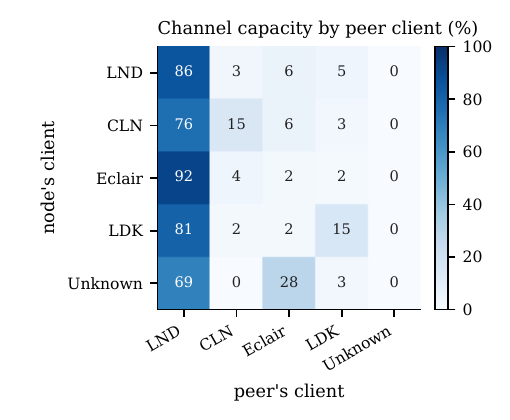}
\caption{Share of each client's channel capacity (rows) by the client of the peer (columns), routable core of
25~August~2026.}
\label{fig:matrix}
\end{figure}

\paragraph*{Dependence on LND, against a null model}
Minority clients' channel capacity faces LND: 76\% for CLN, 92\% for Eclair and 81\% for LDK (Fig.~\ref{fig:matrix}),
and without LND nodes only 32\% of CLN pairs, 51\% of Eclair pairs and 12\% of LDK pairs remain connected for
100k-sat payments (14\%, 51\% and 10\% for 1M sats). Read alone, this suggests that minority clients reach each other
only through LND. A null model says otherwise. Permuting client labels among nodes of similar capacity (within
capacity deciles, 200 draws) keeps the graph and the cohort sizes and removes any preference of a client for its own
kind. Under it, CLN would put 88\% of its capacity on LND peers and LDK 87\%; both put less. CLN nodes
keep 15\% of their capacity among CLN peers against 8\% by chance, and LDK 15\% against 2\%. Without
LND, 9\% of CLN pairs would remain connected by chance (95th percentile 25\%), against 32\% observed; for
Eclair and LDK the observed values lie within the null range. Minority clients' dependence on LND is what their size
implies; CLN operators in fact build more of their own backbone than chance would, but it is too small to stand
alone.

\section{Discussion}
\label{sec:discussion}

\paragraph*{Containment, not collapse}
The first measured client-wide outage in the LN was absorbed. The network as a whole lost about as much
reachability as in an unusual but ordinary two days, and half as much as in the Tor denial-of-service wave a month
earlier, while the affected cohort lost a third of its internal reachability and saw organic force closes double.
Client-correlated failure was costly for the client's own users, not for everyone else, because the nodes that
complied were peripheral.

\paragraph*{Where the risk is}
The blast radius of a correlated failure is set by its overlap with the hubs. A few large nodes, regardless of
client, carry the network's reachability; ACINQ alone outweighs the entire CLN cohort for the survivors. Node-count
diversity statistics therefore mislead in both directions: CLN's 8.5\% of nodes protected little, and LND's 84\%,
four points of which are one operator's fleet, matters mainly through the subset of LND nodes that are hubs. The structure that diversity statistics hide is not a
pathological dependence of minority clients on LND, which is what their size implies, but the concentration of
connectivity in a few operators, whose software choice is a single point of failure for everyone.

\paragraph*{Operational lessons}
Compliance with an emergency call was voluntary and uneven (22\% of CLN nodes, mostly Tor-reachable); a third of
those were still dark after five weeks, so emergency shutdowns have lasting costs for the operators who follow them.
Advising \texttt{--offline} rather than a shutdown was the safer design, because an offline node keeps watching the
chain for revoked states and expiring HTLCs. Single-operator events (mass closes, an exchange migrating its node, a
custodial wallet winding down) are easily mistaken for network-wide effects; concentration checks belong in any
incident analysis. Gossip pipelines should validate dumps (we found saved error responses) and record their vantage
point.

\paragraph*{Recommendations}
(i)~Monitor peer-side disabled shares per client and per transport as an early-warning signal; it is cheap and
was unambiguous for both the Tor and the CLN events. (ii)~Report client diversity weighted by hub capacity or
reachability, not by node count. (iii)~Encourage channels between minority-client hubs, so that CLN, Eclair and LDK
users are not connected only through LND. (iv)~Scope emergency guidance by version and duration, and prefer
\texttt{--offline} to shutdown.

\section{Limitations}
\label{sec:limitations}
Our gossip is one node's view of the announced graph, which changed in mid-2024 and has gaps before 2026.
Unannounced channels (most LDK and Phoenix usage) are invisible; their users reach the network through their
providers' nodes (Phoenix users through ACINQ's), so hub failures cost payments more than \S\ref{sec:resilience}
measures. Feature-bit rules can be wrong for new releases and
are weak where nodes announce no features; Eclair's node count is an upper bound. Node counts treat every node as
an independent operator; we found one managed fleet by its alias and hubs, and smaller fleets without a shared
alias would go undetected. The on-chain signal identifies
wallets, not daemons, and cannot separate CLN from Eclair or identify LDK. Disabled flags record peers' views with
some delay, and a single vantage point can miss or delay updates; such gaps are not specific to one client, and
LND, Eclair and LDK, observed through the same node, did not move on 27~August. A channel between two CLN nodes
that both went dark is never disabled, so $D_{\mathrm{CLN}}$
understates the outage. The dark threshold (80\%) and the two-day footprint are choices; the churn baselines and
placebo windows use the same definitions. The resilience model measures connectivity under static capacity
thresholds, an upper bound on payment success. Channel balances are private, so we do not model how payments
rerouted around failed nodes drain liquidity on the surviving paths, nor fees, HTLC limits or private routing hints.
The absolute reachability levels therefore overstate what payments achieve; the observed outage, the placebo
windows and every scenario are measured with the same model, so the comparisons between them are like for like.
Chain data do not reveal which side broadcast a force close. Event dates are those of public announcements; some
advice circulated earlier in community channels.

\section{Ethics}
We use only public gossip and blockchain data and report aggregates. We name large public operators only for
their closing behaviour, which is public on-chain and in the operators' own announcements, and for ACINQ's widely
known role as a hub; we describe other critical nodes by type and publish no lists of potentially vulnerable nodes.

\section{Conclusion}
\label{sec:conclusion}
LND holds about 85\% of public Lightning capacity, a share that has barely moved in three years. When an emergency
took a fifth of CLN nodes offline in August~2026, the network absorbed it: the damage stayed within the CLN cohort,
and a transport failure a month earlier had cost the network twice as much. The fragility of the network lies not in
how many nodes run each client but in which software its few hubs run. Client diversity protects the Lightning
Network to the extent that it reaches those hubs.

\section*{Acknowledgements}
The author thanks Marisha Sahay for valuable feedback and assistance with data preparation and review.

\section*{Data and Code Availability}
Code, the aggregate datasets behind every figure and table, and a script that recomputes every number quoted in the
text are available at \url{https://github.com/bitcoindatalabs/ln-client-diversity}. All stages truncate their inputs at the data cutoff of
4~October~2026. The underlying gossip and blockchain data are public; our daily gossip snapshots are available from
the author on request.

\bibliographystyle{IEEEtran}
\bibliography{references}

\end{document}